\documentclass[12pt,epsfig]{article}
\usepackage{epsfig}
\newcommand{\beq}{\begin{equation}}   
\newcommand{\eeq}{\end{equation}} 
\newcommand{\bea}{\begin{eqnarray}}   
\newcommand{\eea}{\end{eqnarray}}           
\newcommand{\beqn}{\begin{equation}}   
\newcommand{\eeqn}{\end{equation}}
\newcommand{\mbf}[1]{\mbox{\boldmath $#1$}}
\newcommand{\ba}{k_1^*}
\newcommand{\bb}{k_2^*}
\newcommand{\bh}{\bar{h}}
\newcommand{\bk}{\mbf{k}} 
\newcommand{\bq}{\mbf{q}} 
\newcommand{\bl}{\mbf{l}} 
\newcommand{\al}{\alpha} 
\newcommand{\be}{\beta} 
\newcommand{\ga}{\gamma} 
\newcommand{\as}{\alpha_s}
\newcommand{\asb}{\bar{\alpha}_s} 
\newcommand{\om}{\omega} 

\begin{document}

\vspace*{0.5 cm}

\begin{center}   
\begin{Large}   
\vspace*{0.5cm}  
{\bf On the Triple Pomeron Vertex in Perturbative QCD}\\[1cm]
J.Bartels$^a$,  M.G.Ryskin$^b$, and G.P.Vacca$^c$   
\end{Large}\\[1cm]   
$^a$II. Institut f\"ur Theoretische Physik; Universit\"at Hamburg\\
Luruper Chaussee 149, 22761 Hamburg, Germany \footnote{Supported by the
TMR Network ``QCD and Deep Structure of Elementary Particles''}\\ 
$^b$ St.Petersburg
Nuclear Physics Institute\\ Gatchina, St.Petersburg 188300, 
Russia\footnote{Work supported by the
NATO Collaborated Linkage Grant SA (PST.CLG.976453)5437
and by the RFFI grant 00-15-96610}\\ 
$^c$ Dipartimento di Fisica, Universit\`a di Bologna and \\
Istituto Nazionale di Fisica Nucleare, Sezione di Bologna,\\
via Irnerio 46, 40126 Bologna, Italy
\end{center}   
\vspace*{0.75cm} 
\begin{abstract}
\noindent
We estimate the size of the triple Pomeron vertex in perturbative QCD and 
compare with the phenomenological value extracted from Regge fits to 
experimental data. For simplicity, the results of the QCD analysis are taken 
in the large-$N_c$ limit. We find that the perturbative triple Pomeron 
coupling is of the same order of magnitude as the observed one. 
We also estimate the size of the Pomeron self energy and its contribution 
to the renormalization of the Pomeron intercept. The effect is very small,
in agreement with previous nonperturbative estimates.   
\end{abstract}
\section{Introduction}

The Regge description of hadronic high energy scattering processes contains 
a few fundamental parameters which are of nonperturbative nature. Their 
values have been extracted from the analysis of a large variety of 
experimental data, and, so far, there exist no calculations within QCD 
which would allow a comparison of theory and experiment. 
Prominent examples are the Pomeron intercept $\alpha_P(0)\approx 1.08$ and the 
Pomeron slope $\alpha_P'\approx 0.25$ GeV$^{-2}$, 
seen in the total cross section and in elastic 
scattering, and the triple Pomeron coupling $g_{3P}$, defined and measured in 
high mass diffraction.\\ \\
Whereas the former two parameters refer to the (effective) Pomeron seen
at present energies, it is widely believed that the latter one 
provides information on the origin of the Pomeron: a Pomeron with intercept 
exactly at one would exhibit features that are typical for systems near a 
phase transition point ~\cite{MPT,AB}. 
In such a situation the triple Pomeron vertex which describes the splitting 
of a single Pomeron into two Pomerons then provides the starting point for 
calculating correlation functions, critical indices etc. For example,
using a field theoretic description of the Pomeron, the value of the
triple Pomeron coupling determines the size of the self energy, it 
renomalization of the intercept etc. In reality, the 
intercept is close to unity (but not exactly at one), so it is likely 
that, at present day energies, we are in the vicinity of a phase transition, 
and the triple Pomeron vertex plays a fundamental role.\\ \\       
In perturbative QCD, the Pomeron is approximated by the BFKL calculation 
~\cite{BFKL} (in LO and, more recently, also in NLO ~\cite{FL}). However, the 
values for Pomeron 
intercept and slope are not very close to the observed hadronic values; 
moreover, the BFKL approximation can be justified only for scattering 
processes in which the scattering objects have a small transverse extension 
($\gamma^*-\gamma^*$ scattering, or onium-onium scattering). As to the 
Pomeron slope, for $t\neq0$, the BFKL amplitude predicts a small value,
whereas at $t=0$ the $t$-slope is singular, reflecting thus the long distance 
behavior of the perturbative massless gluons. The next parameter, 
the perturbative triple Pomeron vertex, has first been calculated in  
~\cite{B,BW}, starting form the high energy behavior of 
QCD Feynman diagrams. Later on, independent derivations have been
performed, within Feynman diagrams ~\cite{BV, GPV},
using a Wilson line approach  ~\cite{Balitsky} and within the QCD dipole 
approach ~\cite{M,P}. As far as the numerical computation of 
this perturbative coupling and its comparison with the experimental 
hadronic vertex is concerned, an important step has been 
done in ~\cite{BNP,K}: the analytic expression derived from the underlying 
Feynman diagrams contains conformal integrals which have been computed 
in ~\cite{BNP,K}. These results, however, do not yet allow for a 
direct comparison with experimental data: as it was the case already for the 
BFKL approximation in elastic scattering at $t=0$, also the perturbative 
triple Pomeron vertex has a singularity at zero momentum transfer. 
Any numerical estimate, therefore, will depend upon the way in which this 
singular behavior is treated.\\ \\
There is no doubt that perturbative QCD cannot be used in hadron hadron 
small-angle scattering. Nevertheless, the analysis of perturbation theory 
in this high energy limit provides the first step towards the `real' theory, 
and it is important to see, `how far away from reality' we are in pQCD. 
It is the purpose of this paper, to attempt a numerical estimate of the 
perturbative triple Pomeron vertex and to compare with the hadronic value. 
We start from the Feynman diagram analysis of ~\cite{B,BW}, and we make
use of the numerical values obtained in ~\cite{BNP,K}. 
For reference we use the cross section formula for diffraction in 
the triple Regge region: we derive a value for the perturbative triple 
Pomeron vertex which can be compared to the measured hadronic value.
We also estimate the self energy of the BFKL Pomeron. Some of our results 
differ 
from earlier estimates, contained in the literature ~\cite{BNP,Braun,MS}.\\ \\
When trying to compare BFKL predictions with the Pomeron parameters measured 
in hadron hadron scattering, we will face a few difficulties of general 
nature. First, hadron hadron scattering, to a very good approximation, 
has been parametrized by a simple Regge pole in the complex angular 
momentum plane; the leading BFKL singularity, on the other hand, is a fixed 
cut which leads, in addition to the Regge exponents $s^{\alpha}$, to 
logarithms of the energy. Furthermore, BFKL scattering amplitudes are slightly 
singular when the momentum transfer $t$ is taken to zero; this singularity  
reflects the $1/k^2$-singularity of the zero mass gluon propagator in 
perturbative QCD. Nonperturbative effects, therefore, are expected to 
be particularly strong near $t=0$, and a comparison between 
perturbative Pomeron parameters and the measured nonperturbative values 
looks more promising in the region of nonzero $t$-values.\\ \\
Our paper is organized as follows. In section 2 we briefly review, for 
comparison, the perturbative BFKL Pomeron in elastic scattering. 
In section 3 we turn to the triple Regge region of diffraction and define 
what we mean by a `triple Pomeron vertex' in perturbative QCD.    
Section 4 deals with the self-energy of the BFKL Pomeron.
The numerical evaluation will be done in section 5. In the final section we 
give a summary and a few general comments. Some technical details are
put into two small appendices.    
  
\section{Elastic scattering}

In order to find the correct normalization of the triple Pomeron vertex 
we have to start from elastic $2 \to 2$ scattering. Let us write down the 
Regge ansatz for an elastic $2 \to 2$ scattering 
process. For a Pomeron pole in the complex angular momentum plane the elastic 
amplitude is
\beq
A_{el}=-e^{-i \frac{\pi}{2} \alpha(t)}  g_N^2
s \left( \frac{s}{s_0} \right)^{\alpha(t)-1} \; ,
\eeq
the elastic cross section has the form
\beq
\frac{d \sigma}{d t} = \frac{1}{16 \pi} g_N^4
\left( \frac{s}{s_0} \right)^{2 \alpha(t)-2}, 
\eeq
and the total cross section is
\beq
\sigma^{\rm tot}=  g_N^2 \left( \frac{s}{s_0} \right)^{\alpha(0)-1} \; .
\eeq

In the following we want to compare these expressions with the ones obtained 
from perturbative QCD in the Regge limit. In particular, we have to relate
the residue functions $g_N$ to impact factors which naturally arise in 
a perturbative analysis. For reasons which will become clear soon, we 
will have to define forward and non-forward coupling functions, $g_F$ and 
$G_{NF}$, resp.     

For a $2\to2$ scattering process 
(e.g. gluon-gluon scattering or $\gamma^*-\gamma*$ scattering) a color 
singlet exchange leads, in lowest order $\alpha_s$, to the form    
\beq
A_{el}^{\rm LO}= i \frac{s}{2} 
\int \frac{d^2 \mbf{k}}{(2\pi)^3} \Phi_1(\mbf{k}, \mbf{q}-\mbf{k})
\frac{1}{\mbf{k}^2(\mbf{q}-\mbf{k})^2}
\Phi_2(\mbf{k}, \mbf{q}-\mbf{k})
\eeq
All the momenta are living in the transverse plane, $\mbf{q}$ (with 
$t=-\mbf{q}^2$) denote the 
momentum transfer and $s$ the squared center of mass energy, resp,
and $\Phi_i$  is the impact factor of the scattering particle $i$. 
As an example, with this choice for the integration measure, the gluon impact
factor can be written as $ g^2 2 \sqrt{\pi}\, N_c/\sqrt{N_c^2-1}$.

Summing all the contribution in the leading $\log{s}$ approximation 
leads to the BFKL Pomeron exchange; instead of the two gluon propagators 
we insert the BFKL Greens function. The new amplitude reads
\beq
A_{el}^{LL}=\frac{i s}{2} \int \frac{d^2 \mbf{k}}{(2\pi)^3}
\frac{d^2 \mbf{k}'}{(2\pi)^3}
\Phi_1(\mbf{k}, \mbf{q}-\mbf{k})
G(y| \mbf{k}, \mbf{q}-\mbf{k}; \mbf{k}', \mbf{q}-\mbf{k}')
\Phi_2(\mbf{k}', \mbf{q}-\mbf{k}'),
\label{Ael}
\eeq
where $y$ is the rapidity variable.
Clearly, for $\alpha_s \to 0$, when all rungs of the BFKL resummation
decouple, this expression reduces to the two gluon exchange,
which means that we use the following normalization:
\beq
\lim_{\alpha_s \to 0}
G(y| \mbf{k}, \mbf{q}-\mbf{k}; \mbf{k}', \mbf{q}-\mbf{k}')
=  \frac{(2\pi)^3}{\mbf{k}^2(\mbf{q}-\mbf{k})^2}
\delta^{(2)}(\mbf{k}-\mbf{k}')
\label{Gnorm}
\eeq

\begin{figure}[ht!]
\centering
\includegraphics[width=100mm]{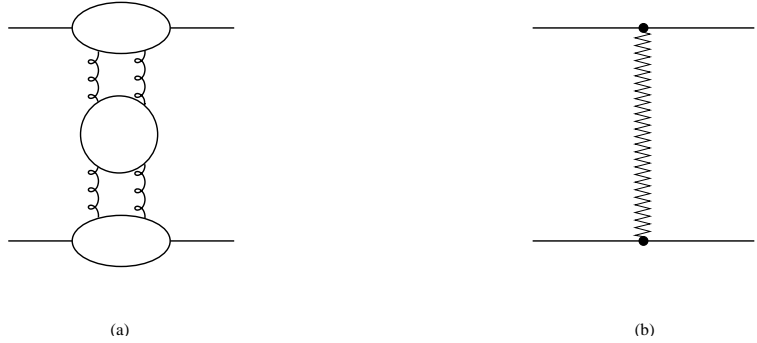}
\caption{Elastic process}
\end{figure}


\subsection{The BFKL Pomeron Green function}

Eigenfunctions of the BFKL kernel, $E_{h,\bh}$, 
are well known in coordinate space, where its form is dictated by
conformal invariance
\beq
E_{h,\bh}(\mbf{r}_{10},\mbf{r}_{20})=
\left(\frac{r_{12}}{r_{10}r_{20}}\right)^h
\left(\frac{r_{12}^*}{r_{10}^* r_{20}^*}\right)^{\bar{h}} \; ,
\label{pom_coord}
\eeq
with $\mbf{r}_{10}=\mbf{r}_1-\mbf{r}_0$ etc,
$h=(1+n)/2+i\nu$,
$\bar{h}=(1-n)/2+i\nu$ ($h^*=1-\bh$, $\bh^*=1-h$), and standard complex
notation for the two-dimensional vector is used on the right-hand side.
Fourier transforming (we use the Lipatov's convention which
assigns a $1/(2\pi)^2$ to any coordinate integration) 
to momentum space one finds \cite{BBCV}
\beq
\tilde{E}_{h\bh}(\mbf{k}_1,\mbf{k}_2)=
\int \frac{d^2\mbf{r}_1}{(2\pi)^2} \frac{d^2\mbf{r}_2}{(2\pi)^2}
E_{h,\bh}(\mbf{r}_1,\mbf{r}_2) 
e^{i (\bk_1 \cdot \mbf{r}_1 + \bk_2 \cdot \mbf{r}_2 )}=  
C\Big(X(\mbf{k}_1,\mbf{k}_2)+
(-1)^nX(\mbf{k}_2,\mbf{k}_1)\Big).
\label{pom_mom}
\eeq
The coefficient $C$ is given by
\beq
C=\frac{(-i)^n}{(4\pi)^2}h\bh (1-h)(1-\bh)\Gamma(1-h)\Gamma(1-\bh).
\label{normpom}
\eeq
The functions $X$ in complex notation can be expressed in terms of 
hypergeometric functions:
\beq
X(\mbf{k}_1,\mbf{k}_2)=\left(\frac{k_1}{2}\right)^{\bh-2}
\left(\frac{\bb}{2}\right)^{h-2}F\left(1-h,2-h;2;-\frac{\ba}{\bb}\right)
F\left(1-\bh,2-\bh;2;-\frac{k_2}{k_1}\right) \; .
\eeq
This analytic form does not contain any term of the type
$\delta^2( \mbf{k}_1)$ or $\delta^2( \mbf{k}_2)$ which are present in the 
coordinate representation (\ref{pom_coord}).
For the impact factor of a colorless external particle we have the 
well known property, that it vanishes for zero gluon momentum: in this 
case the delta-function type contributions do not contribute. For simplicity,
we therefore will ignore them.

The Pomeron intercept has the form $\alpha _{P}(0)=1+\chi(\nu,n)$ 
where 
\beqn
\chi(\nu,n)= 
\bar{\alpha}_s\left( 2\psi(1)-\psi(\frac{1+|n|}{2}+i\nu)
                  -\psi(\frac{1+|n|}{2}-i\nu) \right),\ \
\bar{\alpha}_s=\frac{N_c\alpha_s}{\pi}
\eeqn

Let us now consider the Pomeron Green function in the coordinate
representation \cite{Lcft}. Making use of the Casimir operator properties 
of the M\"obius group, one can chose a representation which is more convenient
to perform a Fourier Transform. In particular, after considering the relation
\bea
\frac{1}{|\rho_{12}|^4} E_{h,\bh}( \rho_{10},\rho_{20})&& =
\frac{1}{16} |\mbf{\partial}_1|^2 |\mbf{\partial}_2|^2   
\frac{1}{(\rho_{12}^2\partial_1 \partial_2) \times (h.c.) } 
E_{h,\bh}( \rho_{10},\rho_{20}) \nonumber \\
&&=\frac{1}{16}\frac{1}{[\nu^2+(n-1)^2/4][\nu^2+(n+1)^2/4]}
 |\mbf{\partial}_1|^2 |\mbf{\partial}_2|^2
E_{h,\bh}( \rho_{10},\rho_{20})
\eea
one can write 
\beq
G_2^{(A)}(y|\rho_1,\rho_2;\rho_{1'}\rho_{2'})= \int d \mu \, d^2\mbf{\rho}_0
\,  e^{y\chi(\nu,n)}
\, N_h \,
  |\mbf{\partial}_1|^2 |\mbf{\partial}_2|^2
E_{h,\bh}( \rho_{10},\rho_{20})
E_{h,\bh}^* ( \rho_{1'0},\rho_{2'0})
\label{G2coord}
\eeq
where we use the measure in the conformal weight space $\int d \mu \, = 
\sum_{n}\int d\nu$ with the following normalization factor $N_h$:
\beq
N_h= \frac{(\nu^2+n^2/4)}{[\nu^2+(n-1)^2/4][\nu^2+(n+1)^2/4]}.  
\eeq

In order to find the momentum representation we take the Fourier transform
taking into account the total momentum conservation, and we obtain:
\beq
\tilde{G}_2^{(A)}(y|\bk_1,\bk_2;\bk_{1'}\bk_{2'})= (2\pi)^3 \int d \mu \,
 e^{y\chi(\nu,n)}
\, N_h \times (2\pi)^2 \,
|\bk_1|^2 |\bq-\bk_1|^2 \tilde{E}_{h,\bh}( \bk_1,\bq-\bk_1)
\tilde{E}_{h,\bh}^*( \bk_{1'},\bq-\bk_{1'}) \;.
\label{G2mom}
\eeq
The $(2\pi)^3$ factor in front of the integral comes from the normalization 
(\ref{Gnorm}); the second $(2 \pi)^2$ factor results from the $\mbf{\rho}_0$
integration, together with a $\delta^{(2)}(\bk_1+\bk_2-\bk_{1'}-\bk_{2'})$
related to the overall momentum conservation. 
As before, $\bq=\bk_1+\bk_2=\bk_{1'}+\bk_{2'}$ is the conserved exchanged 
momentum. This form of the Green function is amputated on the lhs, i.e. 
for the gluons with momenta $\bk_1$ and $\bk_2$.
Clearly, dividing by $|\bk_1|^2 |\bq-\bk_1|^2$  one arrives at the 
nonamputated Green's function $\tilde{G}^{(NA)}_2$.
\subsection{Extraction of the couplings $g_F$ and $g_{NF}$}

Substituting in (\ref{Ael}) the expression of the BFKL Pomeron Green's
function the elastic scattering amplitude can be written as:
\beq
A_{el}^{LL}(\bq)= \frac{is}{2} (2\pi)^5 \sum_n \int d\nu \,  N_h \, 
e^{y\, \chi(\nu,n)} \, \Phi_1^h(\bq) \Phi_2^{h*}(\bq) \; ,
\eeq
where
\beq
\Phi_i^{h}(\bq)= \int \frac{d^2 \bk}{(2\pi)^3}
\tilde{E}_{h,\bh}( \bk,\bq-\bk) \Phi_i( \bk,\bq-\bk)
\label{confimp}
\eeq
are the impact factors in the conformal representation, i.e. 
integrated with the BFKL Pomeron eigenstates, and $\bq$ is the total 
transverse momenta exchanged.

For our purposes it will be sufficient to consider the elastic scattering of 
identical particles. We will consider the forward ($\bq=0$) and the non 
forward ($\bq \neq 0$) case separately.
Since we are interested in the leading high energy behavior,
we restrict ourselves to the conformal spin $n=0$, and we perform the 
integration in $\nu$ in the saddle point approximation for $y \to \infty$.
We therefore need to know the behaviour of $\Phi_i^h$ as a function of $\nu$.
In the appendix we show that the forward ($\Phi_F$) and
non forward ($\Phi_{NF}$) cases are different. Near the saddle point at 
$\nu=0$ we find:
\beq
\Phi_F=\frac{1}{i\nu} \Phi_{0F} +O(\nu^0) \; , \quad \quad
\Phi_{NF}=\Phi_{0NF}+O(\nu) \; ,
\label{relimp}
\eeq
Moreover, we have $N_h=16 \nu^2$. Next we need the expansion 
\beq
\chi_(\nu,0)=\chi_0 -a \, \nu^2 \; , \quad 
\chi_0 = 4 \ln(2) \asb \, \quad a=14 \zeta(3) \asb .
\eeq
In the forward case $\bq=0$ we obtain
\beq
A_{F}= is 8 (2\pi)^5 \frac{\sqrt{2 \pi}}{[2 a y]^{1/2}} \Phi_{0F}^2 \, 
 e^{y\, \chi_0} \; ,
\label{ampliF}
\eeq
while in the non-forward case we have
\beq
A_{NF}= is 8 (2\pi)^5  \frac{\sqrt{2\pi}}{[2 a  y]^{3/2}}
\Phi_{0NF}^2 \, 
 e^{y\, \chi_0} \; .
\eeq
The transition between the non-forward and the forward region is a delicate 
matter. The different large-$y$ behavior ($y^{-1/2}$ and $y^{-3/2}$) 
of the $A_F$ and $A_{NF}$ amplitudes 
originates from the different small-$\nu$ behavior (note, in particular, 
the $1/\nu$ singularity in the forward impact factor (18)). 
As we will show in the Appendix B, this $1/\nu$ singularity
comes from the large-$r_0$ domain, and it reflects the perturbative nature 
of the BFKL Pomeron. In particular, it is related to the singularity of the 
perturbative gluon propagator at zero momentum, and it must disappear after 
the introduction of an appropriate
infrared cutoff \footnote{This was demonstrate using a simplified form
of the BFKL kernel in \cite{CL,DF}.}.
Considering the elastic scattering amplitude as a function of the 
momentum transfer $t$, the difference in the small-$\nu$ behavior of the 
non-forward and the forward results leads to a cusp at $t=0$:
the elastic cross section (2) has a finite limit at $t=0$, but its 
$t$-derivative at $t=0$ is infinite. Generally speaking, in the BFKL 
approximation the point $t=0$ exhibits the perturbative nature most 
explicitly, and changes from perturbative to nonperturbative QCD are expected 
to be most dramatic in this kinematic region.      
 
After these general remarks we are now able to extract, by comparing 
(20) and (21) with (1), the Regge residue factors $g_F$ and $g_{NF}$:
\bea
g_F&=&2^{3/2} \Phi_{0F} (2\pi)^{5/2} \left( \frac{2\pi}{2 a y} \right)^{\frac{1}{4}}
\nonumber \\
g_{NF}&=&2^{3/2} \Phi_{0NF} (2\pi)^{5/2} \left( \frac{2\pi}{(2 a y)^3}
\right)^{\frac{1}{4}}
\label{relcoup} 
\eea
The fact that these couplings have a residual $y$-dependence is a consequence 
of the branch cut nature of the BFKL singularity in the angular momentum 
plane: eqs.(1) -(3) are valid for Regge poles. 
We shall use these relations in the next section in order to extract the 
triple Pomeron vertex $g_{3P}$. 
\section{Triple Pomeron amplitude}

Having collected all necessary ingredients we now turn to the central topic of 
this study, the triple Pomeron vertex. We 
again start from the Regge form for the diffractive cross
section in the triple Regge region, assuming Regge pole singularities in 
all three exchange channels. The cross section is obtained from the 
6-point amplitude (Fig.2) by taking the discontinuity in the diffractive 
mass squared, $M^2$.
We define
\beq
M^2 \frac{d \sigma^{(diff)}}{d t\, d M^2}= \frac{1}{8\pi^2 s} a_6;  
\label{a6def}
\eeq
Regge theory gives
\bea
M^2 \frac{d \sigma^{(diff)}}{d t\, d M^2}
&=& \frac{1}{16 \pi^2}
| e^{-i \frac{\pi}{2}\alpha(t)} |^2 g_N^3 \, g_{3P}
\left( \frac{s}{M^2} \right)^{2 \alpha(t)-2}
\left( \frac{M^2}{s_0} \right)^{\alpha(0)-1} \nonumber \\
&=& \frac{1}{16 \pi^2}
| e^{-i \frac{\pi}{2}\alpha(t)} |^2 g_F g_{NF}^2 \, g_{3P}
\left( \frac{s}{M^2} \right)^{2 \alpha(t)-2}
\left( \frac{M^2}{s_0} \right)^{\alpha(0)-1}
\label{crossdiff}
\eea
where $g_{3P}$ is the triple Pomeron 
vertex, $t$ denotes the momentum transfer, and 
$s_0$ is an energy scale. The triple Pomeron vertex depends upon $t$. 
It will also be convenient to introduce the rapidity 
variable $Y=\log{(s/s_0)}$ and
the rapidity interval of the diffractive states $Y_M=\log{(M^2/s_0)}$.
It will be our aim to extract the counterpart of $g_{3P}$ in the framework
of perturbative QCD (by analyzing, in the leading log approximation, 
the analogous high energy limit in perturbative QCD), and to compare 
its value with the empirical value in $pp$ scattering. To this end we 
consider a (hypothetical) process in the 
triple Regge region, e.g. $\gamma^*\gamma^* \to X \gamma^*$, for which the use 
of perturbative QCD can be justified. Because of Regge factorization, the 
value of the triple Pomeron vertex will be independent of the external 
particles (e.g. $\gamma^*$ with virtuality $Q^2$).        

\subsection{The QCD amplitude}
Let us look at the analysis \cite{BW} of QCD Feynman 
diagrams in the leading $\log s$ approximation and recapitulate the main 
results.
To this end we write the general integral representation for the
differential cross section of the $3\to 3$ process in Fig. 2 
(see eq. ($2.2$) in \cite{BW})

\beq
M^2 \frac{d \sigma^{(diff)}}{d t\, d M^2}= \frac{1}{16 \pi}
\int \frac{d \om}{2\pi i} \int \frac{d \om_1}{2\pi i}
\int \frac{d \om_2}{2\pi i} \left( \frac{M^2}{Q^2}\right)^\om
\left( \frac{s}{M^2}\right)^{\om_1+\om_2} \xi_{\om_1}\xi^*_{\om_2}
F(\om,\om_1,\om_2,0,t,t),
\label{dif1}
\eeq 

where the $\xi_{\om_i}$ are signature factors.

\begin{figure}[ht!]
\centering
\includegraphics[width=60mm]{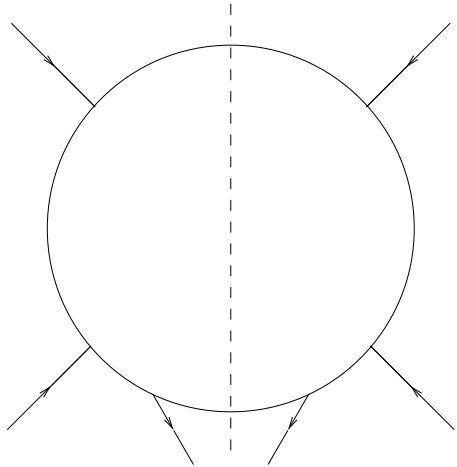}
\caption{Diffractive amplitude}
\end{figure}

The differential cross section (\ref{dif1}) can be represented by
diagrams built from reggeized gluons.
The set of diagrams which is of particular interest for us is illustrated 
in Fig.3a: ellipses denote impact factors, circles the two-gluon 
BFKL Green's function, and the triangle the $2\to4$ gluon vertex
The shaded box denotes the evolution of the 4 gluon state, mediated by the 
sum over all pairwise interactions between the four reggeized gluons.
The last interaction (in Fig.3a: the lowest one) has to connect one of 
the two reggeized gluons on the lhs with one of the gluons on the rhs. Let us 
briefly recapitulate how this result is obtained. The analysis of 
Feynman diagrams in the high energy limit leads to gluon amplitudes 
$D_2$, $D_3$, and $D_4$ which satisfy a set of coupled integral equations
(Fig.4). These functions are nonamputated, i.e. they contain reggeon 
denominators for the outgoing (reggeized) gluon states. Removal (amputation) 
of these reggeon denominators leads to the corresponding functions $C_i$ 
($i=2,3,4$). For example,
\beq
D_4^{(\omega)}= \frac{C_4^{(\omega)}}{\omega -\beta_1 - \beta_2 - 
\beta_3 - \beta_4}.
\label{amputation}
\eeq
In order to obtain the partial wave $F$ of the triple Regge cross section,
we attach two $2\to2$ BFKL Green's functions to the amputated function 
$C_4$, one for the two outgoing gluons on the lhs, another one for the 
two gluons on the rhs. In order to avoid double counting we have to require 
that the last interaction inside the four-gluon state has to connect one of 
the two gluons on the lhs with one of the gluons on the rhs. As a result,
we arrive at the following expression for the partial wave $F$ in the 
triple Regge cross section formula:         
\beq
F= \Biggl[ C_4^{(\om)}-\frac{C_4^{(\om)} \otimes V_{2\to 2}(12)}
{\om -\be_{1'} -\be_{2'}-\be_3-\be_4} -
\frac{C_4^{(\om)} \otimes V_{2\to 2}(34)}
{\om -\be_1 -\be_2-\be_{3'}-\be_{4'}} \Biggr] \otimes G_{2\to 2}(12,\om_1)
G_{2\to 2}(34,\om_2),
\label{Fdef}
\eeq
where $\be_i$ are the trajectories of the reggeized gluons, $V_{2\to
2}$ is the BFKL kernel (without the gluon trajectory function, but including 
its tensor color structure, here acting for the pairs (12) and (34) 
which are in a color singlet state), and $G_{2\to 2}$ 
denotes the full non forward BFKL Green function.
\begin{figure}[ht!]
\centering
\includegraphics[width=150mm]{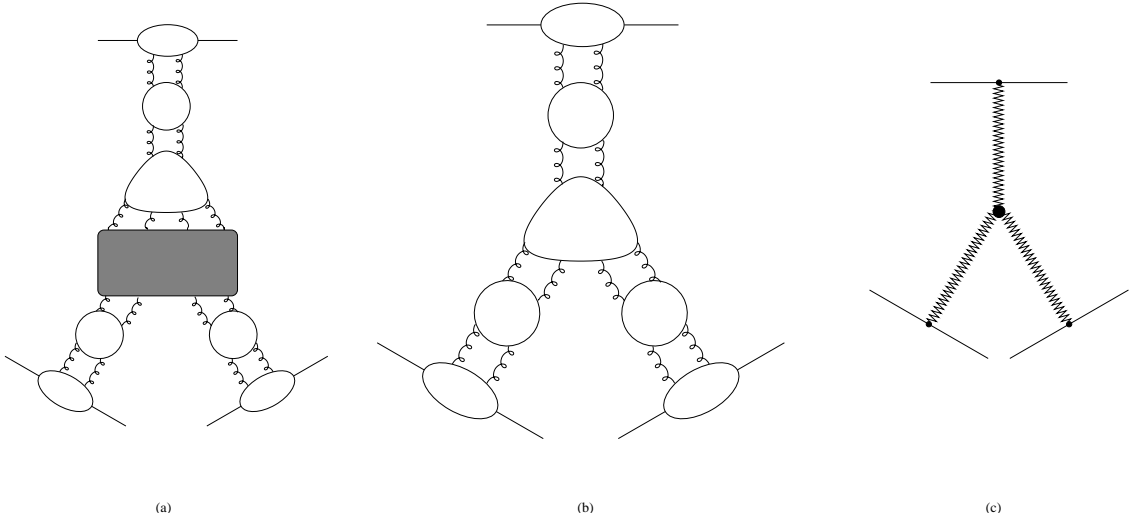}
\caption{Triple interaction}
\end{figure}
A convenient way to rewrite $F$ in terms of $D_4$ is:
\bea
F&=&\left[D_4^{(\om)}\otimes (\om - H_4)+
D_4^{(\om)}\otimes(V(13)+V(14)+V(23)+V(24))\right]\nonumber \\
&&\quad\otimes G_{2\to 2}(12,\om_1) G_{2\to 2}(34,\om_2),
\label{Ffinal}
\eea
where $H_4$ is the standard BKP evolution operator for the
4-gluon state.

As the last step, one has to apply a reduction procedure to the amplitude 
$D_4$. As a part of the coupled integral equations, $D_4$ still contains 
reggeizing pieces: the outgoing four gluon state may contain configurations 
where a pair of two gluons is in an antisymmetric color octet configuration, 
which satisfies the BFKL bootstrap condition and collapses into a single 
gluon. It is convenient to remove these configurations, i.e. to define 
amplitudes $D_4^I$ which are irreducible with respect to the bootstrap 
property. This reduction has been described in ~\cite{BW}, and 
$D_4$ decomposes into the two terms $D_4=D_4^R+ D_4^I$, separating the 
reggeizing (R) and irreducible (I) parts. In (\ref{Ffinal}), let us first 
consider the irreducible part, $D_4^I$. As shown in ~\cite{BW}, $D_4^I$ 
consists of the diagrams of Fig.3a which we have described before.
The triangle - with two gluons entering from above and four gluons 
leaving below - defines the triple Pomeron vertex, and its structure is 
quite simple:
\begin{equation}
\delta_{b b'}
\left( \delta_{a_1a_2}\delta_{a_3a_4} V(12,34)+
 \delta_{a_1a_3}\delta_{a_2a_4} V(13,24)+    
 \delta_{a_1a_4}\delta_{a_2a_2} V(14,23) \right),
\label{vertex2to4} 
\end{equation}
where the $b$, $b'$ are the color labels of the 
reggeized gluons of the ladder above the triple Pomeron vertex, $a_i$
the color indices of the reggeized gluons 
inside the two lower ladders (counting from left to right), and the 
arguments of the function $V$ refer to the momenta of the gluons. 
Below this vertex, before the two gluons on the lhs and the two gluons on the 
rhs are restricted to color singlet states and branch into the two 
disjoint BFKL Green's functions, all pairwise interactions between 
the four gluons have to be summed. However, it is easy to see that any 
rung between two color singlet two-gluon states costs a suppression 
factor of the order $1/N_c^2$: in the large-$N_c$ limit, therefore,
in Fig.3a the interaction inside the shaded area can be neglected, and we 
are left with the diagrams of Fig.3b. In (\ref{Ffinal}), the factor 
$\omega - H_4$ cancels the evolution inside $D_4^R$, and the terms 
proportional to $V(13)$ etc. drop out. In (\ref{vertex2to4}), only the first 
term contributes to the large-$N_c$ limit.       
\begin{figure}[ht!]
\centering
\includegraphics[width=100mm]{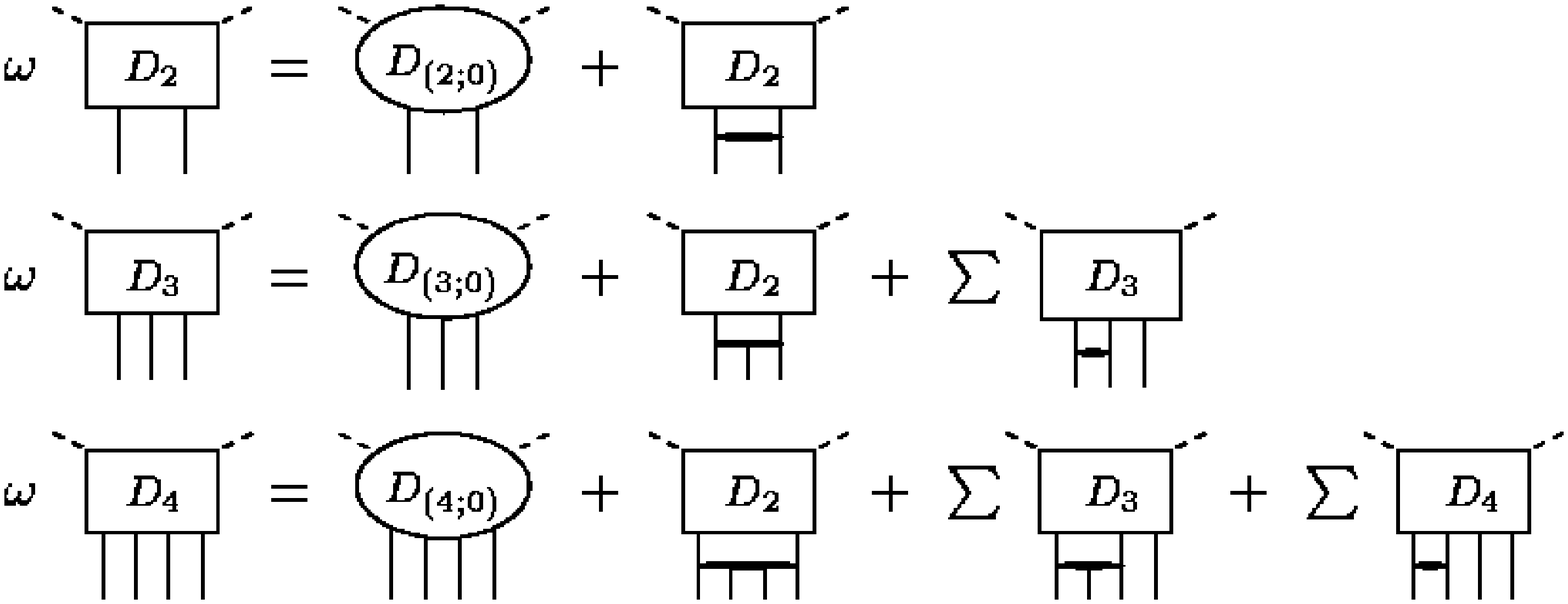}
\caption{Chained equations for the multi reggeized gluon amplitudes.}
\end{figure}

Before we write down the explicit expression for these 
diagrams, a few words about the contribution of $D_4^R$.
It is convenient to go back to (\ref{amputation}) and (\ref{Fdef}). 
Since $D_4^R$ is nothing but a BFKL ladder in which, at the lower end, 
the reggeized gluons split into two (or three) elementary gluons, 
it provides an extra contribution to the triple Pomeron vertex.        
From the color structure of $D_4^R$, given in ~\cite{BW}, eq.(4.3), 
it can be shown that this contribution is subleading in $1/N_c$. 
In conclusion, the large-$N_c$ limit therefore reduces the 
diffractive cross section, given in ~\cite{BW}, to the diagrams shown 
in Fig.3b which is very similar to the `Regge pole' diagrams of Fig.3c.     

\subsection{Extraction of the triple Pomeron vertex}
Let us write down the analytic expression for the 
the generalized six point amplitude $a_6$: 
\bea
 a_6 &&\!\!\!\!\!\!\!\!\! = 2 s \!\!
\int \frac{d^2 \bq}{(2\pi)^3} \frac{d^2 \bq_{1'} d^2 \bq_{2'}}{(2\pi)^3}
\delta^{(2)}(\bl_{\al}-\bq_{1'}-\bq_{2'})\Phi_{\al}(\bq,
\bl_{\al}-\bq)\tilde{G}^{(A)}_2(Y_M|\bq,\bl_{\al}-\bq;\bq_{1'},\bq_{2'})
\nonumber \\
&&
\int \frac{d^2 \bk_1 d^2 \bk_2}{(2\pi)^3} \delta^{(2)}(\bl_{\be}-\bk_1-\bk_2)
\frac{d^2 \bk_3 d^2 \bk_4}{(2\pi)^3}\delta^{(2)}(\bl_{\ga}-\bk_3-\bk_4) 
V_{2\to 4} ( \bq_{1'},\bq_{2'}| \bk_1,\bk_2 ; \bk_3, \bk_4)\nonumber \\
&&
\int \! \frac{d^2 \bk_1'}{(2\pi)^3}
\tilde{G}^{(NA)}_2(Y-Y_M|\bk_1,\bk_2;\bk_{1'},\bl_{\be}-\bk_{1'})
\Phi_{\be}(\bk_{1'}, \bl_{\be}-\bk_{1'}) \nonumber \\
&&\int \! \frac{d^2 \bk_3'}{(2\pi)^3}
\tilde{G}^{(NA)}_2(Y-Y_M|\bk_3,\bk_4;\bk_{3'},\bl_{\ga}-\bk_{3'})
\Phi_{\ga}(\bk_{3'}, \bl_{\ga}-\bk_{3'}),
\label{fullA}
\eea
where $\bl_i$ (with $\bl_i^2-t_i$) are the momentum transfers 
through the impacts factor $\Phi_i$. Later, on we will take $t_{\alpha}=0$ 
and define the triple Regge cross section.

Following the procedure of the previous section, we perform the Fourier 
transform for the $t_{\alpha}$-channel (last
two lines in (\ref{fullA})); we write a representation of 
$\delta^{(2)}(\bl_{\al}-\bq_{1'}-\bq_{2'})$ together with the Green function
in the last line, see(\ref{G2mom}), to obtain
\bea
\label{G2mombis}
\delta^{(2)}(\bl_{\al}-\bq_{1'}-\bq_{2'})&& \!\!\!\!\!\!
\tilde{G}^{(A)}_2(Y_M|\bq,\bl_{\al}-\bq;\bq_{1'},\bq_{2'})
= 
(2\pi)^3 \!\! \int d \mu_0 \, e^{Y_M \, \chi(h_0)}
\, N_{h_0} \times \nonumber \\
&& \!\!\!\!\!\!\!\!\!\!\!\!\!\!\!\!\! \int d^2 \mbf{\rho}_{\al} \,
e^{-i \mbf{\rho}_{\al} \cdot 
(\bl_{\al}-\bq_{1'}-\bq_{2'})} |\bq_{1'}|^2 |\bq_{2'}|^2 
\tilde{E}_{h_{\al},\bh_{\al}}( \bq,\bl_{\al}-\bq)
\tilde{E}_{h_{\al},\bh_{\al}}^*( \bq_{1'},\bq_{2'} ).
\eea
The other two $t-channels$ with their (non amputated) Green functions and 
$\delta$ distribution can be rewritten in a similar
way (using an expansion in the conformal weights $h_{\be}$ and $h_{\ga}$); 
the only difference is the absence of the factors 
$|\bk_{1'}|^2|\bl_{\be}-\bk_{1'}|^2$
and  $|\bk_{3'}|^2|\bl_{\ga}-\bk_{3'}|^2$.
Using the conformal representation of the impact factors $\Phi_i^{h_i}(\bl_i)$
for $ i=\al,\be,\ga$ given in (\ref{confimp}),
we obtain the following expression for (\ref{fullA}):
\bea
a_6 && \!\!\!\!\!\!=2 s
\int d\mu_{\al} d\mu_{\be} d\mu_{\ga} \,
e^{Y_M \, \chi(h_{\al})+ (Y-Y_M)(\chi(h_{\be})+\chi(h_{\ga}) ) } 
N_{h_{\al}} N_{h_{\be}} N_{h_{\ga}}
\Phi_{\al}^{h_{\al}} \Phi_{\be}^{h_{\be}*} \Phi_{\ga}^{h_{\ga}*} 
 \int \prod_{i=\al,\be,\ga} d^2 \mbf{\rho}_i  \times \nonumber \\     
&& e^{i (\mbf{\rho}_{\be} \cdot \bl_{\be}+
\mbf{\rho}_{\ga} \cdot \bl_{\ga} -\mbf{\rho}_{\al} \cdot \bl_{\al}) }
\int \prod_{j=1}^4 d^2 \bk_j  
\tilde{E}_{h_{\be},\bh_{\be}}( \bk_1,\bk_2 ) 
e^{-i \mbf{\rho}_{\be} \cdot (\bk_1+\bk_2)}    
\tilde{E}_{h_{\ga},\bh_{\ga}}( \bk_3,\bk_4 )
e^{-i \mbf{\rho}_{\ga} \cdot (\bk_3+\bk_4)}   \times \nonumber \\          
&&\left( \int d^2 \bq_{1'} d^2 \bq_{2'} \,
V_{2\to 4} ( \bq_{1'},\bq_{2'}|\bk_1,\bk_2 ; \bk_3, \bk_4 )\,
|\bq_{1'}|^2 |\bq_{2'}|^2 \tilde{E}_{h_{\al},\bh_{\al}}^*( \bq_{1'},\bq_{2'} )
e^{i \mbf{\rho}_{\al} \cdot (\bq_{1'}+\bq_{2'})} \right)   
\label{fullA2}
\eea

Before we evaluate the Fourier transform of the last line in (\ref{fullA2}),
we note several simplifications. First,
the $2 \to 4$ vertex will be simplified by the fact that the gluons 
$(1,2)$ and $(3,4)$ couple to two BFKL pomerons in color singlet states.
This fact considerably reduces the number of contributions coming from 
$V_{2\to 4}$: only four identical contributions are left.
Next, we restrict ourselves to the large $N_c$ limit which eliminates 
the nonplanar part of the $2 \to 4$ vertex.
As a result of these simplifications we can write:
\bea
&&\int d^2 \bq_{1'} d^2 \bq_{2'} \,
V_{2\to 4} ( \bq_{1'},\bq_{2'}|\bk_1,\bk_2 ; \bk_3, \bk_4)\,
|\bq_{1'}|^2 |\bq_{2'}|^2 E_{h,\bh}^*( \bq_{1'},\bq_{2'} )  = \nonumber \\
&& C_{1V}
\int d^2 \bq_{1'} d^2 \bq_{2'} \,
A_{2\to 3} (\bk_1,\bk_2 + \bk_3, \bk_4| \bq_{1'},\bq_{2'} )\,
|\bq_{1'}|^2 |\bq_{2'}|^2 \tilde{E}_{h,\bh}^*( \bq_{1'},\bq_{2'} ) 
\eea 
where the constant
\beq
C_{1V}=4 \frac{\pi^{3/2}}{32} g^4 \frac{(2N_c)^2}{\sqrt{N_c^2-1}}=
\frac{\pi^{3/2}}{2} g^4 \frac{N_c^2}{\sqrt{N_c^2-1}}
\approx 2^3  \pi^{7/2} \as^2 N_c\; ,     
\eeq
follows from our choice of the normalizations we have made for the
integration measure and for the impact factors. Details are given in
appendix A. As to the color factors,
we keep only the leading term in the large $N_c$ limit.

After performing the Fourier transform of the last line in
(\ref{fullA2}) \cite{BV,GPV} 
(i.e. performing the transition $\bk_i \to \mbf{\rho}_i$)
one obtains for the triple Pomeron coupling
\footnote{Comparing with an analogous expression in ~\cite{BNP},
we differ in the Casimir operators which are not present in ~\cite{BNP}}   
\bea 
&& \frac{C_{1V}} {(2\pi)^4} \delta^{(2)}(\mbf{\rho}_{23})
\frac{|\rho_{14}|^2}{|\rho_{12}|^2 |\rho_{24}|^2}
 |\mbf{\partial}_1|^2  |\mbf{\partial}_2|^2 
E_{h_{\al},\bh_{\al}}^*( \rho_{1\al},\rho_{4\al} )  =
\nonumber \\
&& \frac{C_{1V}}{ (2\pi)^4} \delta^{(2)}(\mbf{\rho}_{23})          
16 h_{\al} \bh_{\al} (1-h_{\al}) (1-\bh_{\al}) 
\frac{1}{|\rho_{12}|^2 |\rho_{24}|^2 |\rho_{41}|^2} 
 E_{h_{\al},\bh_{\al}}^*( \rho_{1\al},\rho_{4\al} )    
\eea
The Fourier transform of the remaining $\bk_i$ dependent part is 
easily done and leads to the two factors 
$E_{h_{\be},\bh_{\be}}( \rho_{1\be},\rho_{2\be} )$ and
$E_{h_{\ga},\bh_{\ga}}( \rho_{3\ga},\rho_{4\ga} )$.      

As a result, we can write
\bea
a_6 && \!\!\!\!\!\!\!\! = 2 s \,
C_{2V} \int d\mu_{\al} d\mu_{\be} d\mu_{\ga} \,
e^{Y_M \, \chi(h_{\al})+ (Y-Y_M)(\chi(h_{\be})+\chi(h_{\ga}) ) }
N_{h_{\al}} N_{h_{\be}} N_{h_{\ga}}
\Phi_{\al}^{h_{\al}} \Phi_{\be}^{h_{\be}*} \Phi_{\ga}^{h_{\ga}*}
\times \nonumber \\  
&& 16 h_{\al} \bh_{\al} (1-h_{\al}) (1-\bh_{\al}) 
\int \prod_{i=\al,\be,\ga} d^2 \mbf{\rho}_i \,
e^{i (\mbf{\rho}_{\be} \cdot \bl_{\be}+
\mbf{\rho}_{\ga} \cdot \bl_{\ga}-\mbf{\rho}_{\al} \cdot \bl_{\al})  } 
\times \nonumber \\   
&& \int \frac{d^2 \mbf{\rho}_1 d^2 \mbf{\rho}_2  d^2 \mbf{\rho}_4}
{|\rho_{12}|^2 |\rho_{24}|^2 |\rho_{41}|^2}  
E_{h_{\be},\bh_{\be}}( \rho_{1\be},\rho_{2\be} )
E_{h_{\ga},\bh_{\ga}}( \rho_{2\ga},\rho_{4\ga} )
E_{h_{\al},\bh_{\al}}^*( \rho_{4\al},\rho_{1\al} ), 
\label{fullA3}
\eea
where $C_{2V}=C_{1V}/(2\pi)^4$.  

The integral in the last line of (\ref{fullA3}) has been calculated 
explicitly in ~\cite{K,BNP}, where the conformal invariance has been used 
explicitly.  The result can be written in the form
\beq 
\Omega(1-h_{\al},h_{\be},h_{\ga}) \,
\left( \rho_{\al \be}^{-\Delta_{\al \be01}} \rho_{\al \ga}^{-\Delta_{\al \ga}} 
\rho_{\be \ga}^{-\Delta_{\be \ga}} \times (h.c.) \right)
\label{vertexred}
\eeq
where the function $\Omega$ can be found in ~\cite{K,BNP}. 
The exponents are defined for the general conformal
covariant three point function: $\Delta_{ij}=h_i+h_j-h_{k\ne i,j}$ and
remembering to
use for the index $0$ the weight $(1-h_{\al})=\bh_{\al}^*$,
which is due to the fact that one function is complex conjugated.
That means $\Delta_{\al \be}=1-h_{\al}+h_{\be}-h_{\ga}$, etc.                      

We shall now consider the limit $\bl_{\al}=0$ 
(keeping $t_{\beta}=-\mbf{l}_{\be}^2$ 
and $t_{\gamma}=-\mbf{l}_{\gamma}^2$  still independent from each other) 
and perform the remaining 
$\mbf{\rho}$ integrals still present in (\ref{fullA3}). Explicitly we
have to calculate:
\beq
I_c=\int d^2 \mbf{\rho}_{\al} d^2 \mbf{\rho}_{\be} d^2 \mbf{\rho}_{\ga} 
\left( \rho_{\al\be}^{-\Delta_{\al\be}} \rho_{\al\ga}^{-\Delta_{\al\ga}} 
\rho_{\be\ga}^{-\Delta_{\be\ga}} \times (h.c.) \right)
e^{i \mbf{\rho}_{\be} \mbf{l}_{\be} + i \mbf{\rho}_{\ga} \mbf{l}_{\ga} }
\label{Ic}
\eeq
The integration over $\mbf{\rho}_{\al}$ can be done easily:
\beq
\int d^2\rho_{\al} 
\left( \rho_{\al\be}^{-\Delta_{\al\be}} \rho_{\al\ga}^{-\Delta_{\al\ga}}
\times (h.c.) \right)=
f(h_{\al},h_{\be},h_{\ga})
\left( \rho_{\ga\be}^{1-\Delta_{\al\be}-\Delta_{\al\ga}} \times (h.c.) \right) 
\eeq
where
\bea
f(h_{\al},h_{\be},h_{\ga})\!\!\!\!\!\!\! &&= - (-1)^{\Delta_{\al\ga}-\Delta_{\al\be}}
\frac{\Gamma(1-\Delta_{\al\be}) \Gamma(1-\Delta_{\al\ga})}
{\Gamma(2-\Delta_{\al\be}-\Delta_{\al\ga})}
\frac{\Gamma(1-\bar{\Delta}_{\al\be}) \Gamma(1-\bar{\Delta}_{\al\ga})}
{\Gamma(2-\bar{\Delta}_{\al\be}-\bar{\Delta}_{\al\ga})}
\frac{\sin(\pi \Delta_{\al\be}) \sin(\pi \Delta_{\al\ga})}
{\sin(\pi (\Delta_{\al\be}+\Delta_{\al\ga}))}  \nonumber \\
&&= - (-1)^{2(h_{\ga}-h_{\be})} 
\frac{\Gamma(h_{\al}-h_{\be}+h_{\ga})
\Gamma(h_{\al}-h_{\ga}+h_{\be})}{\Gamma(2 h_{\al})}
\frac{\Gamma(\bh_{\al}-\bh_{\be}+\bh_{\ga})
\Gamma(\bh_{\al}-\bh_{\ga}+\bh_{\be})}{\Gamma(2 \bh_{\al})}
 \times \nonumber \\ 
&& \frac{\sin(\pi (h_{\al}-h_{\be}+h_{\ga})) \sin(\pi (h_{\al}-h_{\ga}+h_{\be}))}
{\sin(2 \pi h_{\al}))}
\eea
Therefore one is left with the following integration
\beq
I_c= (-1)^{\Delta_{\be\ga}+\bar{\Delta}_{\be\ga}}
f(h_{\al},h_{\be},h_{\ga}) \int d^2\rho_{\be} d^2\rho_{\ga}  
\left( \rho_{\ga\be}^{1-\Delta_{\al\be}-\Delta_{\al\ga}-\Delta_{\be\ga} }
\times (h.c.) \right)
e^{i \mbf{\rho}_{\be} \mbf{l}_{\be} + i \mbf{\rho}_{\ga} \mbf{l}_{\ga} }.
\eeq
Performing a shift in one of the integration variables, one integration can be 
done to extract the momentum conserving $\delta$ function:
\beq
I_c= (2\pi)^2 \delta^{(2)}(\mbf{l}_{\be}+\mbf{l}_{\ga})
(-1)^{\Delta_{\be\ga}+\bar{\Delta}_{\be\ga}} f(h_{\al},h_{\be},h_{\ga})
\int d^2\rho 
\left( \rho^{1-\Delta_{\al\be}-\Delta_{\al\ga}-\Delta_{\be\ga} }
\times (h.c.) \right)
e^{i \mbf{\rho} \mbf{l}_{\be}}.
\eeq
Also the remaining integral can be done easily. Introducing
$\xi=1-\Delta_{\al\be}-\Delta_{\al\ga}-\Delta_{\be\ga}$ and $\bar{\xi} =
1-\bar{\Delta}_{\al\be}-\bar{\Delta}_{\al\ga}-\bar{\Delta}_{\be\ga}$, we have
\beq
\int d^2\rho  \rho^{\xi} \rho^{*\, \bar{\xi}} e^{i \mbf{\rho} \mbf{l}_{\be}}
= 2\pi \, i^{\bar{\xi}-\xi}\,  2^{1+\xi+\bar{\xi}}\,
\frac{\Gamma(1+\bar{\xi})}{\Gamma(-\xi)}\,\mbf{l}_{\be}^{-\bar{\xi}-1}
\mbf{l}_{\be}^{*\,-\xi-1}
\eeq

Returning to (\ref{Ic})we have
\beq
I_c= (2\pi)^3 \delta^{(2)}(\mbf{l}_{\be}+\mbf{l}_{\ga}) f(h_{\al},h_{\be},h_{\ga})
(-1)^{\Delta_{\be\ga}+\bar{\Delta}_{\be\ga}}
 i^{\bar{\xi}-\xi}\,  2^{1+\xi+\bar{\xi}}\,
\frac{\Gamma(1+\bar{\xi})}{\Gamma(-\xi)}\,\mbf{l}_{\be}^{-\bar{\xi}-1}
\mbf{l}_{\be}^{*\,-\xi-1}
\eeq
with 
\beq
\xi = 1 - \left( (1-h_{\al})+h_{\be}-h_{\ga} \right)
-\left( (1-h_{\al})+h_{\ga}-h_{\be} \right)-
\left( h_{\be} +h_{\ga} -(1-h_{\al}) \right)= h_{\al}-h_{\be}-h_{\ga}
\eeq
and 
\beq
\bar{\xi}= \bh_{\al}-\bh_{\be}-\bh_{\ga} \quad \, 
\xi^* = -1 -\bh_{\al} +\bh_{\be}+\bh_{\ga} \quad \,
\bar{\xi}^* = -1 -h_{\al} +h_{\be} +h_{\ga}
\eeq

Collecting all this in the formula for the amplitude $a_6$ in
(\ref{fullA3}), we obtain the result:
\bea
\!\!\!\!\!\!\!\!\!a_6 \!\!\!\!\!\!\!\!\!\!&& = 2 s \,
C_{3V} \int d\mu_{\al} d\mu_{\be} d\mu_{\ga} \,
e^{Y_M \, \chi(h_{\al})+ (Y-Y_M)(\chi(h_{\be})+\chi(h_{\ga}) ) }
N_{h_{\al}} N_{h_{\be}} N_{h_{\ga}}
\Phi_{\al}^{h_{\al}} \Phi_{\be}^{h_{\be}*} \Phi_{\ga}^{h_{\ga}*}
\times \nonumber \\  
&& \!\!\!\!\! 16 h_{\al} \bh_{\al} (1-h_{\al}) (1-\bh_{\al}) 
\Omega(1-h_{\al},h_{\be},h_{\ga}) \,
 f(h_{\al},h_{\be},h_{\ga}) \times \nonumber \\
&&\!\!\!\!\!(-1)^{\Delta_{\be\ga}+\bar{\Delta}_{\be\ga}}
 i^{\bar{\xi}-\xi}\,  2^{1+\xi+\bar{\xi}}\,
\frac{\Gamma(1+\bar{\xi})}{\Gamma(-\xi)}\,\mbf{l}_{\be}^{-\bar{\xi}-1}
\mbf{l}_{\be}^{*\,-\xi-1}
\label{fullA4}
\eea
Here we have removed the $\delta^{(2)}(\mbf{l}_{\be}+\mbf{l}_{\ga})$ 
function. The constant $C_{3V}$ has the form 
\beq
C_{3V}= (2\pi)^3 C_{2V} = \frac{C_{1V}}{2\pi}  
\eeq

Let us now consider the saddle point approximation, assuming that 
both $Y_M$ and $Y-Y_M$ are large. Following the standard BFKL arguments,
the leading contributions will come from the conformal weights being close 
the value $1/2$, i.e. from conformal spin equal to zero and small $\nu$'s.
In this region we have
\beq
h_i =\frac{1}{2} + i \nu_i, \quad \xi =
-\frac{1}{2} +i (\nu_{\al}-\nu_{\be}-\nu_{\ga}),
\eeq
and therefore
\beq
f(h_{\al},h_{\be},h_{\ga}) \approx - 1 \times
\frac{\Gamma(\frac{1}{2})\Gamma(\frac{1}{2})}{\Gamma(1)}
\times  \frac{\Gamma(\frac{1}{2})\Gamma(\frac{1}{2})}{\Gamma(1)}
\times \frac{1 \times 1}{- \sin{2\pi i \nu_{\al}}}
\approx -\frac{i \pi}{2 \nu_{\al}} \; ,
\eeq
\bea
(-1)^{\Delta_{\be\ga}+\bar{\Delta}_{\be\ga}} i^{\bar{\xi}-\xi}\,  2^{\xi+\bar{\xi}}\,
\frac{\Gamma(1+\bar{\xi})}{\Gamma(-\xi)}\,\mbf{l}_{\be}^{-\bar{\xi}-1}
\mbf{l}_{\be}^{*\,-\xi-1}  \!\!\! &\approx& \!\!\! -1 \times \frac{1}{2}
\frac{\Gamma(\frac{1}{2})}{\Gamma(\frac{1}{2})}
\mbf{l}_{\be}^{-\frac{1}{2} -i(\nu_{\al}-\nu_{\be}-\nu_{\ga})}
\mbf{l}_{\be}^{*\,-\frac{1}{2} -i(\nu_{\al}-\nu_{\be}-\nu_{\ga})} \nonumber \\
 \!\!\!\!\!\!\!
&\approx& \!\!\! -\frac{1}{2} |\mbf{l}_{\be}|^{-1-2i(\nu_{\al}-\nu_{\be}-\nu_{\ga})}
\eea
Moreover, $16 h_{\al} (1-h_{\al}) \bh_{\al} (1-\bh_{\al}) \approx 1$
and $N_h \approx 16 \nu^2$.
We note that the function $f$, near $\bl_\al=\mbf{0}$,
behaves in $\nu_\al$ in the same way as the forward impact factors (18) 
Therefore, with respect to the $t_{\alpha}$ channel we are facing the same 
problem as discussed in the previous section. 
Substituting this behaviour in (\ref{fullA4}) one has
\bea
a_6 \approx && \!\!\!\!\!\!\! 2 s \,
C_{3V} \int d\nu_{\al} d\nu_{\be} d\nu_{\ga} \,
e^{Y_M \, \chi(h_{\al})+ (Y-Y_M)(\chi(h_{\be})+\chi(h_{\ga}) ) }
16^3 \nu_{\al}^2 \nu_{\be}^2 \nu_{\ga}^2
\Phi_{\al}^{h_{\al}} \Phi_{\be}^{h_{\be}*} \Phi_{\ga}^{h_{\ga}*}
\times \nonumber \\  
&& \Omega\left(\frac{1}{2},\frac{1}{2},\frac{1}{2} \right) \,
\frac{i \pi}{4 \nu_{\al}}
|\mbf{l}_{\be}|^{-1-2i(\nu_{\al}-\nu_{\be}-\nu_{\ga})}
\nonumber \\  
= && 2 \, i\, s \, C_{4V} \int d\nu_{\al} d\nu_{\be} d\nu_{\ga}
\, \nu_{\al} \nu_{\be}^2 \nu_{\ga}^2 \,
e^{Y_M \, \chi(h_{\al})+ (Y-Y_M)(\chi(h_{\be})+\chi(h_{\ga}) ) }
\Phi_{\al}^{h_{\al}}  \times \nonumber \\
&& \Phi_{\be}^{h_{\be}*} \Phi_{\ga}^{h_{\ga}*}
|\mbf{l}_{\be}|^{-1-2i(\nu_{\al}-\nu_{\be}-\nu_{\ga})}
\label{saddleA}
\eea
where
\beq
C_{4V}=2^{10} \pi \Omega\left(\frac{1}{2},\frac{1}{2},\frac{1}{2} \right) \,
C_{3V}= 2^9  \Omega\left(\frac{1}{2},\frac{1}{2},\frac{1}{2} \right) C_{1V}
= 2^{12} \pi^{7/2} \as^2 N_c
\Omega\left(\frac{1}{2},\frac{1}{2},\frac{1}{2} \right).
\eeq
The numerical value for $\Omega(\frac{1}{2},\frac{1}{2},\frac{1}{2})$ has been
found in ~\cite{K,BNP}: 
\beq
\Omega\left(\frac{1}{2},\frac{1}{2},\frac{1}{2} \right)= 
2\pi^7 \, _4F_3(\frac{1}{2}) \, _6F_5(\frac{1}{2}) \approx 7766.679
\label{Omega}
\eeq

The remaining part of our saddle point analysis of $a_6$ is analogous 
to what has been done in subsection 2.1:
the integration over $\nu_\al$ goes with an impact factor
in the forward direction, whereas for the two integrations in $\nu_\be$ and 
$\nu_\ga$ we have the freedom to vary $t_{\beta}$ and $t_{\gamma}$: 
following our discussion after (21) we chose to stay away from the 
'most dangerous' points $t_{\beta}=t_{\gamma}=0$, i.e. we perform our 
comparison in the 'safer' region $t_{\beta}=t_{\gamma}\neq 0$. 
Putting $\bl_{\beta}=\bl_{\alpha}=\bl$ and using (18) we find that 
the dependence in $\bl$ in the dominant contribution, as 
selected by the saddle points at $\nu_\al=\nu_\be=\nu_\ga=0$, is just 
$1/|\bl|$. The integrals are trivially done and lead to the result:
\beq
a_6 \approx 2 s \, \frac{C_{4V}}{|\bl|}
\left( \frac{\sqrt{2 \pi}}{[2 a Y_M]^{1/2}} \Phi_{0F} \,
e^{Y_M\, \chi_0} \right)
\left( \frac{\sqrt{2\pi}}{[2 a (Y-Y_M)]^{3/2}} \Phi_{0NF} \,
e^{(Y-Y_M)\, \chi_0} \right)^2 \;.
\eeq
Using the relations in (\ref{relcoup}) for rewriting the impact factors in 
terms of $g_{F}$ and $g_{NF}$, we arrive at:
\beq
a_6 \approx  s \, g_F \, g_{NF}^2
\frac{C_{4V}}{|\bl|} \frac{1}{2^{7/2} \, (2\pi)^{15/2}} 
\left( \frac{2\pi}{2 a Y_M} \right)^\frac{1}{4}  
\left( \frac{2\pi}{[2 a (Y-Y_M)]^3} \right)^\frac{1}{2} 
 e^{Y_M\, \chi_0} e^{2 (Y-Y_M)\, \chi_0} \;.
\eeq

Finally we have to relate the generalized amplitude $a_6$ to the 
triple Regge cross section formula:
\beq
M^2 \frac{d \sigma^{(diff)}}{d t\, d M^2}= \frac{1}{8\pi^2 s} a_6;  
\eeq
Comparing our result with (\ref{crossdiff}), we are able to extract the 
\bea
g_{3P} &\approx& 
\frac{C_{4V}}{|\bl|} \frac{1}{2^{5/2} \, (2\pi)^{15/2}} 
\left( \frac{2\pi}{2 a Y_M} \right)^\frac{1}{4}  
\left( \frac{2\pi}{[2 a (Y-Y_M)]^3} \right)^\frac{1}{2} \nonumber \\
&=& \frac{1}{|\bl|}  \frac{2^6}{(2\pi)^{4}} \as^2 N_c
\Omega\left(\frac{1}{2},\frac{1}{2},\frac{1}{2} \right)
\left( \frac{2\pi}{2 a Y_M} \right)^\frac{1}{4}  
\left( \frac{2\pi}{[2 a (Y-Y_M)]^3} \right)^\frac{1}{2} \;.
\eea
We note that this expression is proportional to $\as^{1/4}$, i.e. 
there is a very mild dependence on the strong coupling. 
Considering $\as \approx 0.3$ we get
\beq
g_{3P} \approx \frac{6.5}{Y_M^{1/4}\;  (Y-Y_M)^{3/2}} \;  \frac{1}{|\bl|} \; .
\eeq
The strong dependence upon the momentum transfer near $\bl=0$ which 
is closely connected with the perturbative zero mass gluon confirms our 
expectation that a comparison between perturbative and nonperturbative 
Regge parameters can be done only in the region of finite momentum transfer,
and one cannot expect more than an order-of-magnitude estimate.

Let us comment on other results of this vertex contained in the literature. 
Within the dipole picture
the triple Pomeron has been derived in ~\cite{M}: an expression for the 
triple Pomeron coupling can be derived from eq.(61), but no explicit 
expression or numerical number $V_0$ has been given in this paper.  
In ~\cite{BNP} explicit expressions for the triple Pomeron vertex can be 
found: our result disagrees,
both in the energy dependence and in the overall normalization.
The result of ~\cite{Braun} is closest to ours, but, again, we disagree in 
the overall normalization and in the energy dependence.     
\section{The Pomeron self energy}

As an important application of the triple Pomeron coupling we estimate 
the size of the Pomeron self energy inside a $2 \to 2$ scattering
amplitude.
To this end we replace, in 
the amplitude $a_6$ of the previous section, the two impact 
factors $\Phi_{\be}^{h_{\be}}$ and  $\Phi_{\ga}^{h_{\ga}}$ by 
another triple Pomeron vertex $V_{2 \to 4}$ which though a BFKL Pomeron 
and another impact factor couples to the lower external particle. 
As a modification of $a_6$, one has to consider an additional momentum 
integration over the Pomeron loop, in the momentum variable $\mbf{l}_{\be}= 
\mbf{l}_{\alpha}-\mbf{l}_{\gamma}$. For simplicity we will restrict ourselves 
to the forward direction $\mbf{l}_{\alpha}=0$. Denoting by $\xi'$ the 
same combination (39) of conformal weights as $\xi$, with
$h_{\al}$ being replaced by $h_{\al'}$, the $\mbf{l}_{\be}$ integral can be 
carried out and leads to the conservation of the conformal weights above 
and below the Pomeron loop: 
\bea
&&\int d^2 \mbf{l}_{\be} \,
\left( \mbf{l}_{\be}^{-\bar{\xi}-1}\mbf{l}_{\be}^{*\,-\xi-1} \right)
\left( \mbf{l}_{\be}^{-\bar{\xi'}-1}\mbf{l}_{\be}^{*\,-\xi'-1} \right)^* = 
\nonumber \\
&&\int d^2 \mbf{l}_{\be}\,
\left( \mbf{l}_{\be}^{-\bh_{\al}+\bh_{\be}+\bh_{\ga}-1} 
\mbf{l}_{\be}^{*\,-h_{\al}+h_{\be}+h_{\ga}-1} \right)
\left( \mbf{l}_{\be}^{*\,h_{\al'}-h_{\be}-h_{\ga}} 
\mbf{l}_{\be}^{\bh_{\al'}-\bh_{\be}-\bh_{\ga}}\right) = \nonumber \\
&&\int d^2 \mbf{l}_{\be}\, \mbf{l}_{\be}^{\bh_{\al'}-\bh_{\al}-1} 
\mbf{l}_{\be}^{*\,h_{\al'}-h_{\al}-1} = 
\frac{(2\pi)^2}{2} \delta_{n_{\al} \, n_{\al'}} \delta(\nu_{\al}-\nu_{\al'})
\label{loop_mom} \;.
\eea
Another important ingredient to the Pomeron self energy is the minus sign 
relative to the BFKL amplitude (5). 

We want to evaluate this loop correction to the elastic scattering amplitude, 
using again the saddle point approximation. Starting from the expression 
(\ref{saddleA}), inserting the result (\ref{loop_mom}) and performing the 
$\nu_{\al'}$ integration, we can write 
\bea
\Delta A_{el}^{LL}&=-i &\frac{(2\pi)^2}{2} \frac{s}{4 (2\pi)^4} \
\left(\frac{C_{4V}}{16(2\pi)^5}\right)^2 \!\!\!
\int dY_1 dY_2  \times \nonumber \\
&&
\int d \nu_{\al} \Phi_{\al}^{h_{\al}}
\nu_{\al}^2 \Phi_{\al}^{h_{\al}*} e^{(Y_1+ Y_2)\, \chi(\nu_{\al})}
\left ( \int d\nu \nu^2   e^{(Y-Y_1- Y_2)\, \chi(\nu)} \right )^2 \; ,
\eea
noting that the $\nu_\be$ and $\nu_\ga$ integrations give identical factors.
Let us also note that all the rapidity intervals must be large enough 
to allow the application of the leading log approximation.
This expression can be compared with a similar result in ~\cite{BNP}.
The remaining $\nu$ integrations give:
\bea
\Delta A_{el}^{LL}&=&-i \frac{s}{2} \,
\left(\frac{C_{4V}}{2^5(2\pi)^6}\right)^2 \Phi_{0F}^2 \!
\int dY_1 dY_2  \, e^{\chi_0 \, (Y_1+ Y_2)}
\left( \frac{2\pi}{2 a (Y_1+Y_2)} \right)^\frac{1}{2} \nonumber \\  
&& e^{2 \chi_0 \, (Y-Y_1- Y_2)}
\frac{2\pi}{[2 a (Y-Y_1-Y_2)]^3}
\label{loop}
\eea

This result represents the one loop self energy correction to the BFKL 
approximation (\ref{ampliF}). Apart from the fractional powers of $\alpha_s$
in front of the rapidity factors, the overall power of $\alpha_s$ inside the 
$C_{4V}$-factors is $\alpha_s^4$: compared to the LL BFKL approximation our 
expression is down by two powers of $\alpha_s$, i.e. the self energy 
correction belongs to NNLO and thus is beyond the NLO corrections calculated 
recently. Note, however, the exponent $2\chi_0$ in the last line: for 
large rapidity intervals $Y-Y_1-Y_2$ this energy factor renders the one-loop 
self energy correction more important than the NLO corrections to the 
BFKL kernel.       

\section{Numerical estimates}
\subsection{Triple Pomeron vertex}

In this final part of our study we use our analytical formulae to obtain  
numerical estimates. We begin with the phenomenological vertex
extracted from the $pp\to p\ +\ X$ data in the framework of the old
triple-Regge analysis \cite{KKPT,FF} and compare with the perturbative 
triple-Pomeron vertex $g_{3P}$.

It was observed that the t-dependence of the triple-Pomeron contribution
to the diffractive dissociation cross section is consistent with the
t-behaviour of the proton-Pomeron vertex square; that is 
$M^2d\sigma /dtdM^2\propto g^2_N(t)$. Hence the t= -{\bf l}$^2$-dependence
of the triple-Pomeron vertex $g_{3P}(t)$ must be small; for the $t$-slope
of $g_{3P}$ we estimate $B_{3P}<1$ GeV$^2$. Therefore we may consider 
relatively large $|{\bf l}|\sim 1-2$ GeV where, in the perturbative
calculation,
we are away from the QCD dangerous region $\bf{l}=0$.
In our normalization the phenomenological analysis gives \cite{KKPT,FF}
\begin{equation}
g_{3P}\sim 0.5 - 1 \, \, {\mbox GeV}^{-1}.
\label{g3pphen}
\end{equation} 
Note that when applying our formula (23) to experimental data and extracting 
a numerical value for the triple-Pomeron vertex, this vertex has to be viewed 
as an {\it effective} vertex, i.e. it already accounts for screening 
corrections due to multi-Pomeron cuts. So the bare vertex may
be larger by a factor of about up to 2 - 4 \footnote{In particular, the gap
survival probability within the ISR energy domain, calculated in the 
formalism of ref. \cite{KMR}, is equal to $S^2=0.25\ -\ 0.33$.}
Thus, at the experiment we "observe" a bare vertex of the order
$g_{3P}\sim 2{\mbox GeV}^{-1}$. 
This value corresponds to the events with a gap size 
$\Delta Y =ln(s/M^2)= Y-Y_M$ between $3$ and $5$.

Turning to the perturbative analysis, we first note that, in order to justify 
the saddle point evaluation of our integrals, we need 
$\Delta Y>4$. This is just the region of $z=y\alpha_s N_c/\pi >1$ where the
asymptotic component of the BFKL solution (with conformal spin $n=0$)
starts to exceed the lowest order two-gluon exchange
contribution. At the same time the width of the saddle point $\delta\nu\sim
1/(a\Delta Y)^2\sim 0.3$ (for $\alpha_s=0.3$) becomes sufficiently small.
Thus, if we choose $Y_M=\Delta Y=Y-Y_M=4$, $\alpha_s=0.3$ and 
${\bf l}=1$ GeV$^{-1}$, 
we obtain $g_{3P}\sim 0.6{\mbox GeV}^{-1}$. It follows from (59) 
and from our discussion before that this value has large theoretical 
uncertainties: changes in $\bf{l}$,
$Y$ and $Y_M$
have a stronger influence on the numerical value of the 
perturbative triple Pomeron vertex than on the nonperturbative triple 
Pomeron coupling. As the result of our analysis, we therefore present 
the range     
\begin{equation} 
g_{3P}\sim 0.2\;\; -\;\; 1.7\;\; {\mbox GeV}^{-1}
\end{equation}
which is related to the ranges of values $3<\Delta Y<5$ and
$0.5 \, {\mbox GeV}^{-1}<{\bf l}<2 \, {\mbox GeV}^{-1}$.
Surprisingly, these numerical value are not far from the experimental 
value discussed before.

Finally, we would like to mention that this perturbative value may be a
little overestimated. Namely, we have to 
remember that, in order to arrive at the triple Regge cross section formula,
we had to take the discontinuity of $a_6$ at $\bl_{\al}=0$. In analogy with 
the impact factor (18), the triple Pomeron vertex at $\bl_{\al}$ has a 
singularity $\sim 1/\nu$ just at the saddle point $\nu=0$; as we have 
discussed in section 2, this singularity is a result of the massless 
gluon propagator, and it should disappear after the introduction of an 
appropriate infrared cutoff. Numerically, the presence of the $1/ \nu$ 
singularity should lead to enhancement, and our value of the perturbative
triple Pomeron vertex may, in fact, therefore be overestimated.

 It is interesting to compare our result with the numerical studies of the
Balitsky-Kovchegov equation \cite{Balitsky,Bal,Kov} (BK) equation. The
equation  includes the LO BFKL evolution and accounts for the triple
Pomeron  coupling summing up the fan diagrams in terms of the
dipole-dipole interaction. In the recent paper \cite{GMS} the
 begining of saturation
 was observed at rather small $Y\sim $ 2 - 3. However first the saturation
 is reached for a large size dipole where the absorptive corrections are
  much
 stronger. This is a dangerous region. Even without the confinement and
 for a fixed $\alpha_s$ coupling we faced here two problems:\\
 on one hand, at small $\bl$ ( $k$ in the notations of \cite{GMS}) the
 $1/\nu$ singularity plays a crucial role , as it was discussed above, \\
 on other hand, the  whole approach can be justified only for the case
 when the rapidity interval $\Delta Y$ occupied by each Pomeron is large
 enough. there is no this condition in BK-equation and the Pomerons can
 split immediately, especially for a large size dipoles.

 To avoid these problems we focus on dipoles of a smaller size, smaller
 than  the initial (input) size $1/k_0$, taken in \cite{GMS}
 to be 1 GeV$^{-1}$. Here the absorptive effects reveal itself   at
 $Y\sim $ 4 - 6 (see Fig.2,4 of ref.\cite{GMS}) . This is in agreement
 with our expectation. Based on the simplified form of the first (order
  of $g_{3P}$) fan diagram contribution (23,24) and taken the efective
  perturbative vertex  $g_{3P}\sim 0.6$ GeV$^{-1}$ (which was evaluated
  just for  $\Delta Y\sim 4$), we found that the triple Pomeron amplitude
  becomes  comparable with the single Pomeron exchange at $Y\sim $ 4 - 5;
  we choose $\bl \sim 1$ GeV, $g_N=10$ GeV$^{-1}$ (corresponding to
  $\sigma^{tot}_{pp}=40$ mb and $\alpha_s=0.2$ (corresponding to
  the LO BFKL $\omega_0=0.56$) as it was done in \cite{GMS}).

\subsection{Renormalization of the Pomeron intercept due to the Pomeron
selfenergy}
Finally, we estimate the size of the Pomeron self energy correction
and its influence on the intercept of the
perturbative BFKL amplitude. In the loop amplitude (\ref{loop}) we still have
the integrals over the rapidities
$Y_{1}$, $Y_{2}$, and the dominant contribution comes from the region
of small $Y_1$, $Y_2$, where $\Delta Y\ =\ |Y - Y_{1}-Y_{2}|\ \to \ Y$.
This limit corresponds to a two-Pomeron exchange of eikonal type.
On the other hand, (\ref{loop}) was derived under the assumption that $Y_1$ and
 $Y_2$ are large
enough to justify the insertion of BFKL Pomerons between the loop and the
impact factors, and we have to restrict $\Delta Y$ to a region smaller
than the total rapidity. If the total rapidity $Y$ is  much larger
than the rapidity interval occupied by the loop - for example we could
consider a loop of the finite size $\Delta  Y$ with
$\Delta Y \sim 1/\omega_0 \sim 4$ -
then such a "small" loops
can be repeated many times and would play the role of the Pomeron
 self energy, leading to a renormalization of the Pomeron intercept.
 As mentioned before,
the self energy is negative relative to the BFKL amplitude, and the
renormalization therefore lowers the Pomeron intercept.
If the absolute value of the renormalization is close or
even larger than the 'bare' intercept $\omega_0=\alpha(0)-1$,
one may get close to the "critical" Pomeron or even obtain the "subcritical"
Pomeron,
as it has been discussed in \cite{MPT,AB,Kaid} (see also \cite{KPT}
where a prescription for the renormalization of the supercritical
(with $\alpha(0)_P > 1$) Pomeron was proposed).

Before we do our estimate, it is useful to recall the nonperturbative
renormalization caused by the pion loop insertion. For a single pion loop
inside the Pomeron (at $t=0$) we have a formula quite analogous to (62).
Instead of $Y_1, Y_2$, convenient variables are:
the size of the loop $\Delta Y$,  the position of the center of the loop
$Y_c=(Y-Y_1+Y_2)/2$. Inside the pion loop we have the transverse momentum
$k_t$ of the pion. Let us fix, for a moment, the value of $Y_c$ and
calculate the pion loop contribution  $\Delta^\pi$:
 \begin{equation}
\Delta^\pi\ =\ \frac{3g_\pi^2}{16\pi^3}\int d(\Delta Y)dk^2_t
\frac{[k^2_0e^{-\Delta Y}]^2}{[k^2_t+m^2_\pi +k^2_0e^{-\Delta Y}]^2 } \
\approx\  \frac{3g_\pi^2 k^2_0}{16\pi^3}\int d(\Delta Y) e^{-\Delta Y} \
\sim\  0.1\ .
\label{loop_pions}
\end{equation}
Here the first factor 3  results from the contributions of
$\pi^+$, $\pi^0$ and $\pi^-$. For the numerical estimate
we neglect the pion mass $m_\pi$ , put the Pomeron pion coupling
square $g^2_\pi=\sigma_{\pi\pi}(s_0)\sim 20$ mb and  choose the mean
transverse momentum of the first particle inside the Pomeron-pion
vertex $k_0\sim 0.6$ GeV. One needs this particle also to fix the
rapidity of the vertex $Y_1$ (or $Y_2$).
The integral over $\Delta Y$ (\ref{loop_pions})
is convergent, and for $\Delta Y << Y$
the integration over $Y_c$ gives a factor $Y$. As a result,
the one loop correction to the amplitude is equal to $Y\cdot \Delta^\pi$.
Inserting two pion loops we obtain $\frac 12 (Y\cdot \Delta^\pi)^2$,
and summing over an arbitrary of loops we get the sum
$exp(Y\cdot \Delta^\pi)$ which means
that the Pomeron intercept increases by $\Delta^\pi$.

Returning to the perturbative Pomeron loop insertion (62), we again choose
$\alpha_s=0.3$ and assume
$Y>> \Delta Y$. Dividing the amplitude (\ref{loop}) by the elastic forward
amplitude (20) and considering only the integration over the loop size,
 we find:
\begin{equation}
\Delta^{loop}\ =\ \int\frac{2\pi\ e^{\chi_0 \Delta Y}}{[a \Delta Y]^3}
\frac{\cdot C^2_{4V}}{(4\pi)^{17}}d(\Delta Y)\ \approx\
 2.53\cdot 10^{-3}\frac{\delta Y}{[\Delta Y]^3} e^{\omega_0\Delta Y} \  .
\label{loop_3pp}
\end{equation}
Here we have cut the integration over the loop size at $\delta Y = \Delta Y$.
As the numerical coefficient here is extremely small, the only
possibility to obtain a relatively large renormalization
is to chose a very large $\Delta Y >> 1/\omega_0$.
On the other hand the loop renormalization $\Delta^{loop}$
is a NNLO BFKL effect and first we have to account for the
NLO BFKL corrections which lowers the intercept down to
$\omega_0 \sim 1/4$. Therefore up to a very  large $\Delta Y$
the loop renormalization is still negligible;
for $\omega_0=0.25$ the values of  $\Delta^{loop}$
and $\omega_0$ become comparable only for $\Delta Y > $ 45
(changing the value of $\omega_0$ to the larger value $\omega_0=0.70$ we
are consistent with ~\cite{MS,Salam}).
Of course, from the academic point of view we have to account for this
renormalization effect, when $s\to\infty$;  but at any reachable rapidity
interval the value of $\Delta^{loop}$ is much less than $\omega_0$.

Since the numerical value of the triple Pomeron
coupling is not far from the nonperturbative one, it is not surprising
to see that the estimate (\ref{loop_3pp}) is also consistent with the
nonperturbative evaluation. In the latter case we expect
\begin{equation}
\Delta^{loop}_{n.p.}\ \sim \ \frac{g_{3P}^2}{16\pi^3}\int d(\Delta Y)
dk^2_t e^{\omega_0\Delta Y}\ \approx\ \
2\cdot 10^{-3}\delta Ye^{\omega_0\Delta Y}\
\label{loop_3pnp}
\end{equation}
for $g_{3P}\sim 1$ GeV$^{-1}$ and $k^2_t\sim 1$ GeV$^2$.

Finally we mention that the estimate in (66) may still be a bit too large.
Namely, recall that
in our calculation of the Pomeron self energy and in our numerical
estimate
of its magnitude  we have restricted ourselves to the forward direction,
that we have considered the values $\bl_\alpha=\bl'_\alpha=0$.
As discussed before, this is the point where the perturbative nature of
the BFKL approximation becomes most visible, i.e. the 'distance' between
pQCD and nonperturbative QCD is the largest. The mathematical
manifestation
is the $1/ \nu$ singularity, which immediately disappears if we
depart from $\bl_{\al}=0$. Therefore, as for the discussion of the
 numerical
value of the triple Pomeron coupling, we expect that also our estimate of
the renormalization due to the selfenergy may be slightly overestimated.

 \section{Conclusions}
In this paper we have performed an estimate of the perturbative triple
Pomeron vertex. Starting from the results of a leading-$\ln s$ analysis
of QCD perturbation theory in the triple Regge limit, we have used the
large-$N_c$ limit to derive a fairly simple expression for the triple
Regge inclusive cross section, which can be compared with the standard
formulae used in the analysis of experimental data. A numerical estimate
of the perturbative triple Pomeron coupling - which has a considerable
theoretical uncertainty - indicates that its value is of the same
order of magnitude as the nonperturbative one, obtained from
earlier fits to experimental data.

We have also tried to estimate the renormalization of the BFKL intercept
due the Pomeron self energy loop. Formally speaking, this is NNLO effect
and lies beyond the NLO corrections to the BFKL kernel. A numerical
estimate
-  again with theoretical uncertainties - indicates that these
corrections due to self interactions of the BFKL Pomeron are
much smaller than the nonperturbative contribution (\ref{loop_pions}).
This is caused
by the fact that this correction is proportional to the
perturbative triple Pomeron vertex square, and it agrees with the
evaluation of the Pomeron loop insertion based on
the phenomenological value (\ref{g3pphen}) of the triple Pomeron vertex..

So finally we conclude that in spite of a `huge number' $\Omega =7767$
(see eq.(\ref{Omega})) the perturbative triple Pomeron coupling is not large;
it is in approximate agreement with the old phenomenological evaluations. 

As a future step, it would be interesting to understand better why
numerical studies of the nonlinear evolution equations
seem to find rather rapid saturation effects in spite of a rather
small value of the perturbative triple Pomeron
coupling.

\section *{Appendix A: Counting factors of $2$ and $\pi$}
In this appendix we give a brief summary of the normalization of the impact 
factors and the triple Pomeron vertex in perturbative QCD. 
Our starting point is eq.(4) which defines the impact factor.
To be definite we consider the elastic scattering of two quarks (averaged over 
color and helicity of the incoming quarks). The lowest 
order diagram with color singlet exchange has two gluons in the t-channel
(box diagram and its crossed counterpart), and in the high energy limit     
on finds
\begin{equation}
A_{el}^{LO}=is g^4 \frac{N_c^2-1}{(2N_c)^2} \int \frac{d^2\mbf{k}}{(2\pi)^2}
\frac{1}{\mbf{k}^2 (\mbf{q}-\mbf{k})^2}.
\end{equation}
Comparison with (4) yields the quark impact factor
\begin{equation}  
\Phi_q = 2 \sqrt{\pi} g^2 \frac{\sqrt{N_c^2-1}}{2N_c}.
\end{equation}
Equivalently, we could have defined the impact factor through
the requirement that the leading order energy discontinuity 
should have the form 
\begin{equation}
disc\;\; A_{el} = \int \frac{d^2\mbf{k}}{(2\pi)^3} \Phi_q 
\frac{1}{\mbf{k}^2 (\mbf{q}-\mbf{k})^2}  \Phi_q.
\end{equation}  

Next we turn to the triple Regge cross section. Eq.(23) defines 
the $M^2$-discontinuity of a six-point function, $a_6$. Turning again to 
lowest order QCD diagrams, we look at the diffractive process 
$q+q \to (qg)+q$a in the triple Regge limit; a typical QCD diagram is shown 
in Fig.5. We define the triple Pomeron vertex through the LO equation
\begin{eqnarray}
a_6 = 2s \int \frac{d^2\mbf{q}}{(2\pi)^3} \frac{d^2\mbf{k_1}}{(2\pi)^3}
\frac{d^2\mbf{k_3}}{(2\pi)^3} \Phi_q \frac{1}{(\mbf{q}^2)^2} 
V(\mbf{q},-\mbf{q}|\mbf{k}_1,\mbf{l}-\mbf{k}_1;\mbf{k}_3,\mbf{l}-\mbf{k}_3)
\nonumber \\
\frac{1}{\mbf{k}_1^2} \frac{1}{(\mbf{l}-\mbf{k}_1)^2} \Phi_q 
\frac{1}{\mbf{k}_3^2} \frac{1}{(\mbf{l}+\mbf{k}_3)^2} \Phi_q.
\end{eqnarray}
The analysis of the high energy behavior of the relevant QCD diagrams leads 
to the result:
\begin{equation}
4 a_6 = \pi^3 \frac{N_c^2-1}{2N_c} g^{10} \int \frac{d^2\mbf{q}}{(2\pi)^3} 
          \frac{1}{(\mbf{q}^2)^2} 
\tilde{V}(\mbf{q},-\mbf{q}|\mbf{k},\mbf{l}-\mbf{k};\mbf{k},\mbf{l}-\mbf{k}) 
          \frac{1}{\mbf{k}_1^2} \frac{1}{(\mbf{l}-\mbf{k}_1)^2} 
          \frac{1}{\mbf{k}_3^2} \frac{1}{(\mbf{l}+\mbf{k}_3)^2}  
\end{equation}     
where $\tilde{V}$ stands for the BFKL-type $2 \to 4$ gluon vertex: 
\begin{equation}
\tilde{V}=\frac{(\mbf{q}^2)^2}{(\mbf{q}-\mbf{k}_1)^2
(\mbf{l}+\mbf{k}_3-\mbf{q})^2}
- \frac{\mbf{q}^2}{(\mbf{q}-\mbf{k}_1)^2 \mbf{k}_1^2}
- \frac{\mbf{q}^2}{(\mbf{l}+\mbf{k}_3)^2 (\mbf{l}+\mbf{k}_3-\mbf{q})^2}
\label{Vmomdef}
\end{equation}
Inserting the result for the quark impact factor we obtain 
the following normalization of the triple Pomeron vertex $V$:
\begin{equation} 
V = \frac{(2N_c)^2 g^4}{\sqrt{N_c^2-1}} \frac{\pi^{3/2}}{32} \tilde{V}
\label{normV}
\end{equation}

\section*{Appendix B: Impact factors in the conformal approximation}

We discuss here briefly, with simple arguments and some approximations, the behaviour
of the impact factors in conformal representation in the forward and non forward
direction, but without giving a full momentum and conformal weight dependence.

Let us consider already the situation of zero conformal spin and work in coordinate
representation. One can obtain the same results on studying the limiting case
of the BFKL Pomeron eigenstate in momentum representation.

We start from an impact factor which have some dominant support in a bounded region
of size of order $R$ in the coordinate space.
We are interested in studying the behaviour of the following expression
\beq
\Phi(\nu,\bl)=\int \frac{d^2\mbf{r}_1}{(2\pi)^2} \frac{d^2\mbf{r}_2}{(2\pi)^2}
|r_{12}|^{1+2i\nu} \Phi(\mbf{r}_1,\mbf{r}_2)
\int \frac{d^2\mbf{r}_0}{(2\pi)^2} e^{i \mbf{r}_0 \bl}
(|r_{10}| |r_{20}|)^{-1-2i\nu} \; .
\label{impbeh}
\eeq
We are mainly interested in studying the behaviour in the small $|\bl|$ region.
Therefore the main contribution in (\ref{impbeh}) will come from the integration
in the region of large  $|\mbf{r}_0|$. It is therefore convenient to split the
integration region according to $|\mbf{r}_0|<R$ and $|\mbf{r}_0|>R$.
We shall be interested therefore in momenta $|\bl|< 1/R$ and neglect the first
contribution. We note also the in the region $\mbf{r}_0|>R$ it is a good
approximation to consider $|r_{i0}|\approx |\mbf{r}_0|$ for $i=1,2$ since
the external integral has support roughly for $|r_{i}|<R/2$. 
We can therefore write, in a factorized form, 
\beq
\Phi(\nu,\bl) \approx 
\int \frac{d^2\mbf{r}_1}{(2\pi)^2} \frac{d^2\mbf{r}_2}{(2\pi)^2}
|r_{12}|^{1+2i\nu} \Phi(\mbf{r}_1,\mbf{r}_2)
\int_{|\mbf{r}_0|>R} \frac{d^2\mbf{r}_0}{(2\pi)^2} e^{i \mbf{r}_0 \bl}
(|r_0|^2)^{-1-2i\nu} = \phi_\nu \;  g(\nu,\bl) \,
\eeq
and study the $\bl$ dependence in $g(\nu,\bl)$.
The $\mbf{r}_0$ integration gives
\bea
g(\nu,\bl)&=&\pi R^{-4 i \nu} \frac{1}{2 i \nu} \left(
\,_1F_2(-2i\nu, 1, 1-2i\nu; - \frac{R^2 \bl^2}{4})
- \frac{\Gamma(1-2i\nu)}{\Gamma(1+2i\nu)}
\left(\frac{R^2 \bl^2}{4}\right)^{2i\nu} \right)
\nonumber \\
&=& \pi R^{-4 i \nu} \left\{ \frac{1}{2 i \nu} \left[
1- \left(\frac{R^2 \bl^2}{4}\right)^{2i\nu} \right] +\frac{R^2 \bl^2}{4} \right\}+O(\nu)+
O\left(\left(\frac{R^2 \bl^2}{4}\right)^2\right)
\eea
One can see that in the forward direction $\bl=\mbf{0}$ the second term does not give
contribution and therefore $g(\nu,\bl)\sim 1/(2i\nu)$. This leads infact to the
correct behaviour of the BFKL Pomeron Green's function in forward direction, where
under the $\nu$ integration the integrand is not proportional to $\nu^2$, as,
instead, in the non forward case.
To analyze the limit of small $\bl R^2 \to \mbf{0})$ one can keep the term
$(R^2 \bl^2)^{2i\nu}/(2i\nu)$ and estimate with the saddle point method
its contribution in such a limit. Again one can easily check that such a contribution
is suppressed.
For $\bl \ne \mbf{0}$ there is instead a cancellation of the $\nu$ pole in the
origin. Therefore we can write
\bea
\Phi(\nu,\mbf{0}) && \approx \frac{1}{i \nu} \Phi_{0F} \nonumber \\
\Phi(\nu,\bl) && \approx \Phi_{0NF} \; , \quad \quad s \gg \bl^2 >0
\label{impsad}
\eea 

We shall be interested typically, for the non forward case, to values of
$R|\bl| \sim 1$, at the border of the approximations taken above to show the
behaviour in (\ref{impsad}).  

\end{document}